\documentclass[aps,a4paper,12pt,superscriptaddress,onecolumn]{revtex4}

\usepackage[T1,T2A]{fontenc}
\usepackage{graphicx}
\usepackage{amssymb,amsmath}
\usepackage{amsfonts}
\usepackage[hypertexnames=false]{hyperref}
\usepackage{enumitem}
\setitemize{topsep=0pt,parsep=0pt,itemsep=0pt}
\newcommand{\ILNP}{Laboratory of Nuclear Problems, Joint Institute for Nuclear Research, RU-141980 Dubna, Russia}
\newcommand{\IAlm}{Institute of Nuclear Physics, KZ-050032 Almaty, Kazakhstan}
\newcommand{\IAst}{L.N.~Gumilyov Eurasian National University, KZ-010000 Astana, Kazakhstan}

\begin{document}

\title{Centrality criteria of inelastic nucleon-nucleon collisions}
\author{V.I.~Komarov}\email{komarov@jinr.ru}\affiliation{\ILNP}
\author{B.~Baimurzinova}\affiliation{\ILNP}\affiliation{\IAlm}\affiliation{\IAst}
\author{A.~Kunsafina}\affiliation{\ILNP}\affiliation{\IAlm}\affiliation{\IAst}
\author{D.~Tsirkov}\email{cyrkov@jinr.ru}\affiliation{\ILNP}

\begin{abstract}
\vspace{0.2cm}

A process of inelastic nucleon-nucleon collisions with the impact parameter less than a radius of the nucleon quark core is considered. The criteria for selection of such collisions from a full array of the nucleon-nucleon interactions are suggested.
\end{abstract}

\maketitle

\label{sec:intro}
\section{Introduction}
The investigation of a non-perturbative QCD (NPQCD) structure of baryons is a fundamental challenge of strong interaction physics. Such a structure guarantees the very existence of the nucleon stable states and determines cardinal properties of the nuclear matter. The main experimental basis for clarification of the NPQCD structure of baryons is the baryon spectroscopy and the short-range nucleon-nucleon interaction. The maximum sensitivity of the nucleon-nucleon data in this aspect can be obtained in the interaction conditions which allow achieving the overlap of the quark core of the nucleons and ensuring its sufficiently long existence. The first condition can be met in $NN$-collisions with an extremely small impact parameter $R$: $R < r_{\mathrm{core}} \approx 0.4$~fm. Collisions of this type will hereinafter be called central collisions. A geometrical probability $k$ of such collisions at the GeV energies is rather high: $k = \pi r^2 / \sigma_{NN}^{\mathrm{total}} = 0.50$~fm$^2/40$~mb$~=0.125$, but the dominant background caused by peripheral collisions conceals central collisions from detailed investigation. Therefore, the criteria for selection of the events of interest from the whole array of the collisions assume great importance.

The centrality criteria used in the studies of heavy nucleus collisions are based on measurement of multiplicity of the particles produced in the collision. The well-known information on the nucleon structure of the colliding nuclei and rather well studied properties of the interaction participants is used~\cite{Broniowski, Das}. Similar information is absent in the case of the nucleon-nucleon collisions, and low multiplicity of the collision products at relatively low energies depends primarily on the energy and does not allow definite evaluation of the centrality. Therefore, special criteria are needed in the case of the nucleon-nucleon collisions.

It should be recalled that division of impacts into central and peripheral was established as far back as 1960s. The most definite results were obtained in measurements of the proton emission in the inelastic inclusive process $pp \rightarrow pX$ at the CMS energies 4.5--7.6~GeV~\cite{Anderson}. It was shown that spectra of protons emitted at angles larger than $29^{\circ}$ with momenta above 1~GeV/$c$ (CMS) well follow the phase space distributions for the system of a proton pair and a certain number of pions. This statistical picture is very different from the picture of emission of protons at small angles, which is mainly determined by a peripheral process of single and double baryon excitation. According to Anderson-Collins estimation~\cite{AndersonCollins} these central collisions make up $(34 \pm 13)\%$ of the full number of the collisions. Their analysis showed a significant role of the nonperipheral impacts in the inelastic proton-proton collisions but did not lead to establishing a relation between the centrality and the impact parameter of the collision and specifying the centrality criteria.

The aim of the present paper is to specify these criteria. The energy conditions necessary for effective interaction of nucleon cores are also considered.

\label{sec:scenario}
\section{A feasible scenario of central collisions}
The nucleon central region overlap may occur only at a sufficiently high energy of collisions. Indeed, if the kinetic energy $W$ of the colliding nucleons is less than a value of the repulsive potential $U_{\mathrm{rep}}(0)$ of the $NN$-interaction at the zero distance between the nucleon centers, that is, $W =\sqrt{s} - 2m_N < U_{\mathrm{rep}}(0)$, then the nucleon wave functions cannot overlap. This circumstance determines the minimum energy $\sqrt{s}_{\mathrm{min}}$ for the overlap to occur
\begin{equation}\label{eq2}
\sqrt{s}_{\mathrm{min}} = U_{\mathrm{rep}}(0) + 2m_N.
\end{equation}
Only a few unrelated values are known for the $U_{\mathrm{rep}}(0)$, which were obtained for different $NN$ states and in different theoretical models: 1.2--2.3~GeV~\cite{Maltman}; 0.83--1.36~GeV~\cite{Stancu}; 0.87--1.45~GeV~\cite{Bartz}; 0.49--0.60~GeV~\cite{Ishii}. Considering large spread of these values, only their characteristic interval can be evaluated at a level of 0.5--1~GeV. This corresponds to the minimal energy $\sqrt{s}_{\mathrm{min}} \approx$ 2.4--2.9~GeV, above which it makes sense to consider the reliable overlap of the nucleon quark content. Strictly central collisions at lower energies are the elastic scattering at $180^{\circ}$ or excitation of single or both nucleons leaving them as separate three-quark clusters that decay with emission of two nucleons and mesons. At energies higher than $\sqrt{s}_{\mathrm{min}}$ the initial six-quark system takes the form of an ensemble of states of two separate three-quark clusters and a state of a joint six-quark system controlled by the QCD symmetries. Overcoming of the repulsion between approaching three-quark clusters means a transition from a meson-baryon state to a six-quark state with the corresponding degrees of freedom. There are two radically different collision channels: (a) the elastic channel, in which three quarks of one nucleon are elastically scattered from three quarks of the other nucleon with the structure of both nucleons fully preserved and (b) the inelastic channel, in which at least one of the quark-quark scatterings has the inelastic character and leads to considerable transformation of the nucleon pair wave function. In the latter case, there appears an intermediate excited six-quark system, a quark bag $(6q)^*$, the study of which may provide new information about the non-perturbative QCD structure of nucleons.
The Hamiltonian of the $(6q)^*$ state in a potential approach can be expressed as a sum
\begin{equation}\label{eq3}
H = \sum_{i=1}^6 (m_q)_i + \sum_{i=1}^6 T_i - T_G + \sum_{j > i=1}^6 V_{ij},
\end{equation}
where $(m_q)_i$ denotes the mass of the $i$-th quark, $T_i$ and $T_G$ are the kinetic energy operators of the $i$-th quark and the center-of-mass motion respectively, and $V_{ij}$ are the potentials of the interaction between the quarks $i$ and $j$. In the center-of-mass system, neglecting the interaction between the quarks, we obtain the maximal momentum $p_q$ of a single quark:
\begin{equation}\label{eq4}
p_q = (s/36 - m_q^2)^{1/2}.
\end{equation}
For the threshold energy of 2.9~GeV, with the mass of the constituent $u,d$ quark taken to be $m_q = 0.34$~GeV, we get $p_q = 0.34$~GeV/$c$. This value exceeds the characteristic value $\Lambda_{\mathrm{conf}}$ = 0.1--0.3~GeV of the momentum corresponding to the transition from a hadronic to a quark state. Thus, at the energies above the threshold value $\sqrt{s}_{\mathrm{min}}$ a six-quark system determined by the QCD degrees of freedom can be produced in the central collisions.

A substantially important factor is the quark momentum region of a quark momentum occurring at this transition. As far back as the 1980s, it was realized~\cite{Shuryak, Manohar} that quark systems with quark momenta lower than $\Lambda_{\chi\mathrm{SB}}$ at which spontaneous breaking of the chiral symmetry occurs, exist in a specific state in which quarks are not the current ones with masses of about 2--5~MeV, but the constituent quarks with above-mentioned significantly higher masses; the interaction between them is through exchange of gluons at short distances ($< 0.1$~fm), Goldstone bosons of the $\pi$, $K$, $\eta$ octet at intermediate distances (0.1--0.5~fm), and through the confinement effect at long distances (0.5--1.0~fm). This interaction is supposed to be much more intensive than in a perturbative quark-gluon system. In some conditions it can even produce quasi-bound states, resonance dibaryons. Therefore, the system expands and decays more slowly than a quark-gluon perturbative QCD system ($t_{\mathrm{decay}} > 0.4$~fm/c$~= 1.3 \cdot10^{-24}$~s) and has time to establish an intermediate state. This state may have a stochastic character or acquire a definite structure corresponding to the QCD symmetries. The mass spectrum of the states includes not only continuum but also eigenvalues of the quasi-bound states. Use of the central $NN$ collisions for searching for resonances of this kind was recently proposed in\cite{Komarov18}. In any case, the state $(6q)^*$ under consideration should be treated as a six-quark chiral constituent state. A special character of the emerging state is caused by the high baryon and energetic density of its matter since this matter with the baryon number $B = 2$ and CMS energy $\sqrt{s}$ is concentrated in a volume of about $4/3 \pi r_{\mathrm{core}}^3$.

The experimental value of $\Lambda_{\chi\mathrm{SB}}$ has not been found yet. Theoretical evaluations are at the level of 1.2~GeV/$c$~\cite{Manohar} and 0.9~GeV/$c$~\cite{Melnitchouk}. In approximations made earlier for equation~\eqref{eq4} with $\Lambda_{\chi\mathrm{SB}} = 1.2$~GeV/$c$ one may evaluate the highest energy at which the chiral constituent quark regime still takes place
 \begin{equation}\label{eq5}
\sqrt{s}_{\mathrm{max}} \approx 6(\Lambda_{\chi\mathrm{SB}}^2+ m_q^2)^{1/2}= 7.5~\text{GeV.}
\end{equation}
This value is significantly underestimated since the main part of the collision energy is spent on the meson field generation. At considerably higher energies constituent quarks are disrupted to produce current quarks, which interact much weaker and cannot create any quasi-stable intermediate system.

Decay of the intermediate $(6q)^*$ system arising at energies of the central collision in the interval
\begin{equation}\label{eq6}
2.9~\text{GeV} \lesssim \sqrt{s} \lesssim 7.5~\text{GeV,}
\end{equation}
leads to reconstruction of hadronic states in the form
\begin{equation}\label{eq7}
p + p \rightarrow (6q)^* \rightarrow N + N + \mathfrak{M},
\end{equation}
where at energies below the antibaryon production threshold $\sqrt{s}_{\mathrm{anti}} = 3.8$~GeV, $\mathfrak{M}$ denotes the system of light mesons, predominantly pions from direct production or decay of other light mesons $\eta, \sigma, \rho, \omega, \varphi$. A certain part of the mesons should be kaons of pair-production (less intensive channel of hyperon production is excluded from the consideration here). The system $\mathfrak{M}$ remains predominantly mesonic and at energies above $\sqrt{s}_{\mathrm{anti}}$ since the antibaryon yield is significantly less than that of mesons. The final states of a similar type are also produced in the dominant peripheral collisions, so that the use of the centrality criteria is a necessary condition for the study of the central collisions.
\label{sec:criteria}
\section{Criteria for identification of central \texorpdfstring{$\boldsymbol{NN}$}{NN} collisions}
Central $pp$ collisions in the elastic channel are distinguished by the scattering at the angle of $90^{\circ}$. The energy dependence of this scattering is well described by the regularity of the ``constituent counting rule'' (CCR)~\cite{Matveev, Brodsky}. The observed small deviations from the CCR~\cite{Schremp, Pire} seem to indicate some influence of the non-perturbative QCD structure in these conditions and has not attracted due attention yet.
A dominant channel of central collisions is the inelastic channel. It is well seen from ratio of the differential cross section for the elastic $pp$ scattering at $90^{\circ}$ to the total cross section $\pi(r_{\mathrm{core}})^2$ of the central collision divided by $4\pi$. The cross section $d\sigma/d\Omega (90^{\circ})$ quickly drops with energy as $s^{-10}$ in accordance with the CCR, so that this ratio is 0.2 at $\sqrt{s} = 2.9$~GeV and is $5\cdot10^{-7}$ at $\sqrt{s} = 6$~GeV. For this reason, it is right the properties of the excited intermediate state $(6q)^*$ that are of a primary interest for gaining information on the NPQCD structure of the $NN$ interaction.

The $(6q)^*$ system is formed via mutual deceleration of the colliding nucleons. This process involves elastic and inelastic rescatterings of the valence quarks of the initial state. These rescatterings suppress the longitudinal component of the initial momentum $p_0 = (s/4 - m_N^2)^{1/2}$. The kinetic energy corresponding to this component dissipates into the energy of the initial nucleon structure transformation, the kinetic energy corresponding to the acquired transversal momentum components, and the energy of the meson field excitation. This process is the most intensive for the strictly central impact where the quarks of the intermediate state lose the initial dominance of the longitudinal momentum components via the mutual deceleration. Accordingly, the nucleons formed at the transition of the $(6q)^*$ bag to the hadron state lose the longitudinal momentum component and are emitted with a significant probability at angles close to $90^{\circ}$. At collision with a high impact parameter the quarks of the initial nucleon far from the collision axis do not exhibit rescattering with the quarks of the counter-nucleon and retain on average their longitudinal momentum component $1/3 p_0$. Accordingly, the final state nucleons acquire a longitudinal momentum and move along the collision axis in opposite directions with the momentum being the higher the larger the impact parameter. Thus, it follows from these obvious considerations that for distinguishing central collisions one should select events with nucleons of a small longitudinal momentum, that is, nucleons emitted at angles close to $90^{\circ}$ CMS. Identification of inelastic central collisions becomes most effective if one selects emission of both nucleons at such angles. In this case, elastic central collisions are automatically excluded, and contribution of inelastic peripheral collisions producing nucleons scattered along the axis of impact in the opposite direction is minimized. At the same time, the relative momentum of the final nucleons is minimized, which leads to strong interaction between them. This interaction results in formation of the bound $^{3}\!S_1-{}^{3}\!D_1$ state, the deuteron, for an isoscalar nucleon pair, and the quasi-bound $^{1}\!S_0$ state, the $S$-wave diproton $\{pp\}\!_S$, for an isovector pair. From here on, for the sake of brevity, these pairs will be referred to as joined pairs. From the aforesaid it follows that in order to identify central collisions, one should select the joined pairs emitted at angles close to $90^{\circ}$ CMS.

Thus, the first criterion for selection of central collisions is detection of reactions like
\begin{subequations}\label{eq8}
\begin{align}
N + N& \to d(90^{\circ}) + \mathfrak{M}, \label{eq8a}\\
N + N& \to \{pp\}\!_s(90^{\circ}) + \mathfrak{M}. \label{eq8b}
\end{align}
\end{subequations}

The main distinction between these reactions is difference of the isospin states of the produced meson systems which affects the dynamics of the joined pair production but does not influence the kinematics of the processes.

\begin{figure}[htbp]\centering
\includegraphics[width=0.7\textwidth]{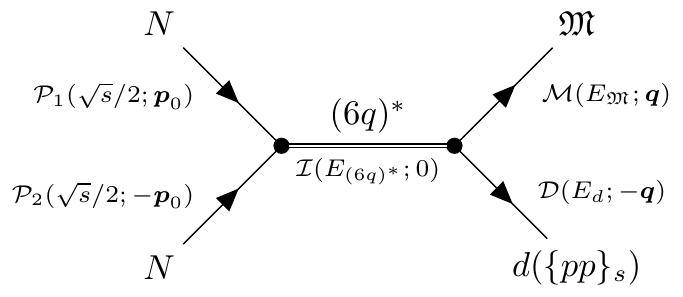}
\caption{The diagram of a central collision with excitation of the intermediate six-quark system in the $s$-channel of the reaction resulting in production of a joined nucleon pair and a meson system.}
\label{diagram}
\end{figure}

The second criterion is smallness of the interaction region size
\begin{equation}\label{eq9}
r_{\mathrm{int}} < r_{\mathrm{core}}.
\end{equation}
A size $r_{\mathrm{int}}$ of the interaction region may be evaluated via the momentum transfer $\mathcal{Q}$ determining the dynamics of the process, $r_{\mathrm{int}} \approx 1/\mathcal{Q}$. The reaction under consideration is a transition of the initial state nucleons with a high relative momentum to the final state joined nucleon pair via excitation of an intermediate state system in the $s$-channel of the reaction (see Fig.~\ref{diagram}).

Each of the initial nucleons undergoes transition to a state of one of the joined pair nucleons. Therefore, the 4-momentum $\mathcal{Q}$ determining the process is $\mathcal{Q} = \mathcal{P}_1 - \mathcal{D}/2$, where $\mathcal{P}_1$ is the momentum of one of the initial nucleons, $\mathcal{D}$ is the momentum of the final joined pair, and $r_{\mathrm{int}} = 1/(- \mathcal{Q}^2)^{1/2}$. It is easy to find that
\begin{equation}\label{eq10}
(-\mathcal{Q}^2)^{1/2} = \frac{1}{2}\left\{-\left[\sqrt{s}-(m_d^2 + q^2)^{1/2}\right]^2 + \left[s - (2m_N)^2 + q^2\right]\right\}^{1/2},
\end{equation}
where
\begin{equation}\label{eq11}
q = \frac{1}{2\sqrt{s}}\left\{\left[s-(m_d + m_\mathfrak{M})^2\right]\left[s-(m_d - m_\mathfrak{M})^2\right]\right\}^{1/2}.
\end{equation}

\begin{figure}[htbp]\centering
\includegraphics[width=0.7\textwidth]{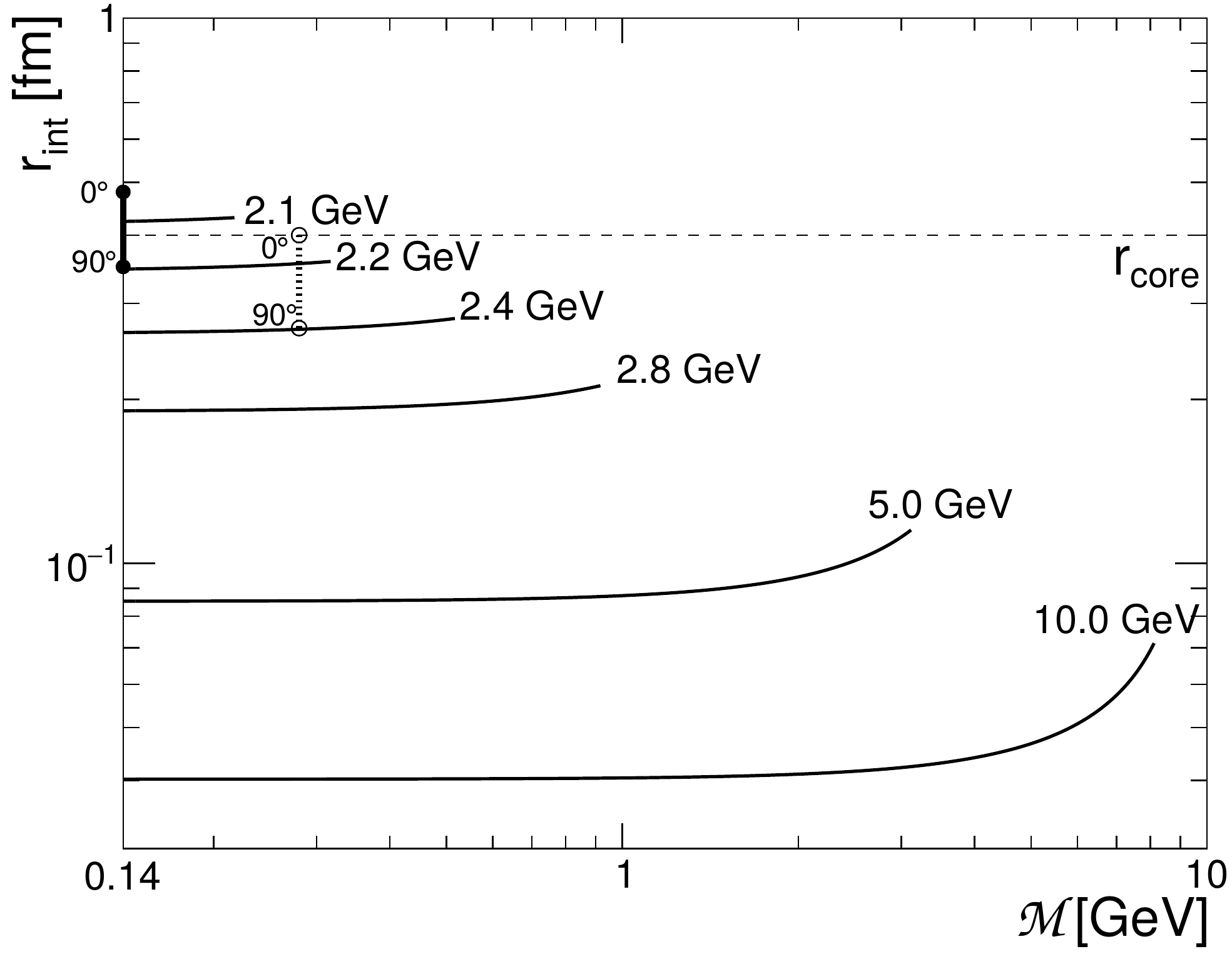}
\caption{Dependence of the interaction region size on the invariant mass of the produced meson system and the collision energy. $\bullet$: $pp\rightarrow \{pp\}\!_s\pi^0$, $\sqrt{s} = 2.2$~GeV; $\circ$: $pn \rightarrow d \pi^0 \pi^0$, $\sqrt{s} = 2.38$~GeV. $0^{\circ}$ and $90^{\circ}$ are the angles of the joined pair emission.}
\label{r_M}
\end{figure}

Here $m_N$ is the nucleon mass, $m_d$ is the deuteron ($\{pp\}\!_s$) mass, and $m_\mathfrak{M}$ is the invariant mass of the meson system.
It is seen from formulas~\eqref{eq10},~\eqref{eq11} that the size of the interaction region $r_{\mathrm{int}} = (\mathcal{Q}^2)^{1/2}$ is uniquely defined by of the invariant mass $m_\mathfrak{M}$ of the produced meson system and the collision energy $\sqrt{s}$. Figure~\ref{r_M} shows that the events with $m_\mathfrak{M}$ larger than the pion mass satisfy centrality criterion~(\ref{eq9}) at energies $\sqrt{s}$ higher than 2.15~GeV. In the same figure the $r_{\mathrm{int}}$ values are shown as an example for the reaction $pp\rightarrow\{pp\}\!_s\pi^0$ at $\sqrt{s} = 2.2$~GeV~\cite{Komarov16} and $pn\rightarrow d \pi^0\pi^0$ at $\sqrt{s} = 2.38$~GeV~\cite{Adlarson}. The momentum transferred depends on the angle of the emitted joined pair. It is seen that at the emerging angle of $90^{\circ}$ the collision proceeds in the region of the nucleon core overlapping.

\begin{figure}[htbp]\centering
\includegraphics[width=0.7\textwidth]{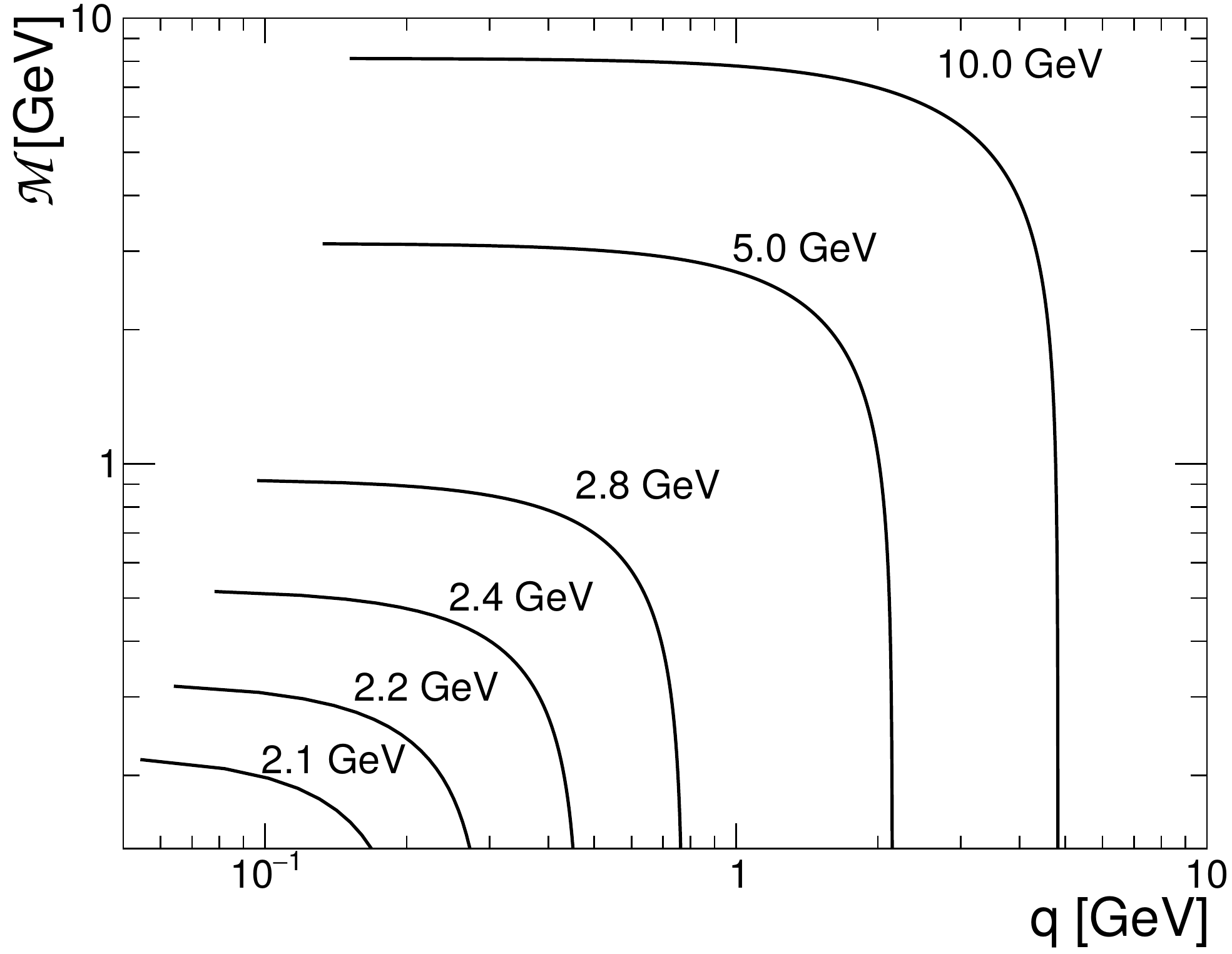}
\caption{Kinematical dependence of the invariant mass of the produced meson system on the momentum of the joined pair, $d$, $\{pp\}_S$.}
\label{M_q}
\end{figure}

Variation of $m_\mathfrak{M}$ from the minimum value $m_{\mathrm{min}} = m_{\pi}$ to the maximum $m_{\mathrm{max}} = \sqrt{s} - m_d$ changes the momentum $q$ of the joined pair from the maximum value $q_{\mathrm{max}}$ defined by formula~\eqref{eq11}, down to zero. It provides the possibility of measuring the spectrum of the meson system mass by measuring the momentum distribution of the joined pair (Fig.~\ref{M_q}).

Practical feasibility of the proposed centrality criteria can be evaluated as follows. Taking the total cross section of the central collisions to be $\sigma_{\mathrm{centr}} = \pi r_{\mathrm{core}}^2$ and assuming isotropy of the proton emission, one gets the differential cross section of their emission at the angle of $90^{\circ}$ $d\sigma_{\mathrm{centr}}(90^{\circ}) /d\Omega_p = r_{\mathrm{core}}^2/4 = 5$~mb/sr. Numerous experimental data (see, e.g.,~\cite{Abramov, Diddens}) show that at energies of 2--20~GeV the deuteron yield relative to the proton one is about $3\cdot10^{-3}$ for a wide range of processes and emission conditions. Therefore, the differential cross section of the deuteron emission is $d\sigma_{\mathrm{centr}}(90^{\circ}) /d\Omega_d \approx 15$~$\mu$b/sr. Taking the solid angle of registration $\Omega = 2$~sr and the luminosity $L = 10^{30}$ cm$^{-2}$s$^{-1}$ as realistic values for typical experimental setups at proton accelerators at GeV energies, one obtains the event registration rate of about 30~s$^{-1}$. It means the possibility of acquiring a rather large amount of information about the processes of interest in a reasonable time. Therefore, central nucleon collisions can become an effective test-bench for exploration of the non-perturbative QCD structure of nucleons.

\label{sec:conclusion}
\section{Conclusion}
Consideration of the feasible scenario of the central $NN$ collisions allows criteria to be formulated for the experimental identification of the collision events leading to overlap of the central quark volumes of the colliding nucleons. Study of characteristics of these events may provide a new perspective source of information for clarification of the non-perturbative QCD structure of the nucleons. The relevant experiments have not been conducted yet, and consideration of their goals and the information that might be extracted from them is a currently important task. The approach to this task is a topic of the next paper.

The authors are grateful to A.V.\,Kulikov and V.I.\,Kukulin for their interest in the problem and useful remarks.

\bibliographystyle{apsrev}
\bibliography{criteria}
\end{document}